# Increased material differentiation through multi-contrast x-ray imaging: a preliminary evaluation of potential applications to the detection of threat materials

A. Astolfo[1,2], I.G. Haig[2], D. Bate[2,1], A. Olivo[1], P. Modregger[3,4]

1. Department of Medical Physics and Biomedical Engineering, UCL, London WC1E 6BT, UK
2. Nikon X-Tek Systems Ltd, Tring, Herts, HP23 4JX, UK
3. Department of Physics, University of Siegen, 57072 Siegen, Germany
4. Center for X-ray and Nano Science CXNS, Deutsches Elektronen-Synchrotron DESY, 22607 Hamburg, Germany

**Abstract**
Most material discrimination in security inspections is based on dual-energy x-ray imaging, which enables the determination of a material's effective atomic number ($Z_{eff}$) as well as electron density and its consequent classification as organic or inorganic. Recently phase-based "dark-field" x-ray imaging approaches have emerged that are sensitive to complementary features of a material, namely its unresolved microstructure. It can therefore be speculated that their inclusion in the security-based imaging could enhance material discrimination, for example of materials with similar electron densities and $Z_{eff}$ but different microstructures. In this paper, we present a preliminary evaluation of the advantages that such a combination could bear. Utilising an energy-resolved detector for a phase-based dark-field technique provides dual-energy attenuation and dark-field images simultaneously. In addition, since we use a method based on attenuating x-ray masks to generate the dark-field images, a fifth (attenuation) image at a much higher photon energy is obtained by exploiting the x-rays transmitted through the highly absorbing mask septa. In a first test, a threat material is imaged against a non-threat one, and we show how their discrimination based on maximising their relative contrast through linear combinations of two and five imaging channels leads to an improvement in the latter case. We then present a second example to show how the method can be extended to discrimination against more than one non-threat material, obtaining similar results. Albeit admittedly preliminary, these results indicate that significant margins of improvement in material discrimination are available by including additional x-ray contrasts in the scanning process.

**Introduction**
Security inspections at e.g. airports are based on dual-energy x-ray imaging methods [1,2]. Images created at two significantly different (average) x-ray energies can be processed with established algorithms [3,4] in an attempt to determine the electron density and the effective atomic number ($Z_{eff}$) of the scanned material. Subsequent research looked into the possibility to use more than two energies [5,6], typically demonstrating better material determination or reduced uncertainty.
As a completely independent line of research, phase-based x-ray imaging, gained momentum in the mid-90s [7-9], following pioneering developments in the mid-60s [10]. Alongside the ability to detect phase changes, access to an additional "contrast channel" was demonstrated in the early 00s [11-13], which was termed dark-field or "Ultra-Small Angle X-Ray Scatter" (USAXS) imaging. This contrast channel is related to the degree of inhomogeneity that the imaged object presents on a scale smaller than the spatial resolution, and indeed this signal was later on connected to "traditional" small-angle x-ray scatter [14-15].

Technology was then developed that enables translating initially the phase-based methods [16,17], then also the dark-field capabilities [18-19] for use with conventional, laboratory-based x-ray sources, which made the technology more widely available.

The research presented in this paper combines all of the above through the use of a scanner based on one of the existing laboratory-based phase technologies ("edge illumination", EI), which uses apertured masks to generate phase sensitivity [17,19-21]. Thanks to the use of a detector with energy-thresholding capabilities [22], the scanner is capable of delivering five contrast channels (attenuation at three different energies and dark-field at two) through a single object scan. More specifically, the detector threshold allows splitting the used x-ray spectrum in two resulting in the collection of high and low energy attenuation ($Abs_H$, $Abs_L$) and dark-field ($Scatt_H$, $Scatt_L$) images; in addition to this, the small percentage of x-rays transmitted through the mask septa are also collected, resulting in the creation of a fifth attenuation image at a much higher average X-ray energy (which we refer to as "offset" image, as it corresponds to the detected intensity between two consecutive beamlets formed by the apertures, i.e. the offset above which the beamlet intensity is detected). The system is also capable of simultaneously registering differential phase at two energies [23,24], but this property is not exploited in this study.

Alongside the established methods that exist to combine attenuation-based images at different energies [3-6], recently approaches have emerged that address dual-energy dark field imaging in a quantitative manner [25]. This paper follows a more basic, simplified approach in which the detection of a material of interest (e.g. an explosive) is maximised against other materials by producing a linear combination of the various contrast channels with floating coefficient, and selecting the set of coefficients that results in the maximum contrast-to-noise ratio (*CNR*). This is done both on $Abs_H$, $Abs_L$ only, as a surrogate for conventional dual-energy imaging, and with the full set of five contrasts ($Abs_H$, $Abs_L$, Offest, $Scatt_H$, $Scatt_L$). Despite the simplicity of the approach, the comparison of the optimised *CNR* in the two cases provides an estimate of the detection advantages that can be obtained by simultaneously exploiting five contrasts instead of two. After laying out the procedure to distinguish two materials from each other and presenting a practical example, we outline an approach that can be used to maximise the detection of a material of interest against multiple others. Although in both cases we provide examples in a security context, the proposed approach is general, and can be applied to the discrimination of any type of materials.

**Materials and methods**

A schematic of the imaging system is shown in Fig. 1. It features a tungsten X-Tek (Tring, UK) 160 x-ray tube with an approximately 80 micron focal spot, operated at 80 kVp and 2 mA. The detector was a CdTe CMOS-based photon counter XC-Flite FX2 manufactured by Direct Conversion. It has 100 micron square pixels and an overall field of view of 20 cm (vertical) times 1.28 cm (horizontal). The detector features two thresholds, one of which is used to cut off the noise, and the other to split the spectrum in two. This was calibrated at the beginning of the experiment by comparing experimental measurements with a theoretical model.

The masks were fabricated to the authors' design by Creatv Microtech (Rockville, MD), by electroplating a ~200 micron thick gold layer on a patterned graphite substrate. Pre-sample and detector masks were placed at 1.50 and 1.95 m from the source, respectively, with the detector placed immediately downstream of the detector mask; their overall size matches the detector's once magnification is taken into account. Aperture sizes were 28 μm and 21.4 μm for detector and pre-sample mask, respectively.

While a symmetric mask is shown for simplicity in Fig. 1, in truth the system employs the "asymmetric" mask concept [26] that enables the acquisition of all image frames necessary for the retrieval of attenuation, differential phase and dark-field images in a single object scan. Both masks are mounted on motor stacks that enables their alignment with each other and with the detector's pixel columns; a third, longer translation stage is used to scan the objects through the beam, simulating the use of a conveyor beld in e.g. an airport scanner for carry-on baggage..

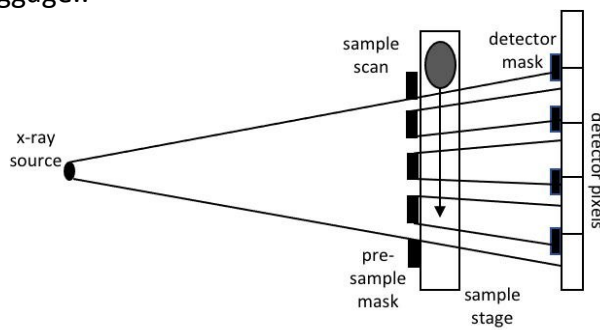

Fig 1 schematic diagram of the used imaging system

Scans with the sample present are acquired alongside "air" scans, and the intensity, central position and full-width at half maximum (FWHM) of the beamlets are compared on a pixel-by-pixel basis. More specifically, beamlets are fitted with Gaussian curves in both cases, at which point the pixelwise ratio between curve areas provides the sample's attenuation, and the difference between curve centres and FWHMs the refraction and dark-field signals, respectively. In the latter two cases, division by the sample-to-detector distance enables converting beamlet shifts/broadenings on the detector into angular values; a full equation-based description is not repeated here for simplicity's sake, and the reader is referred to recent publications [23,27].

Two phantoms simulating explosive concealment in a postal delivery were created to demonstrate the technology in a security-related application. The first one, aimed at developing and testing the approach, was deliberately simpler. It consisted of a thin (2 cm) plastic box containing Semtex H1 placed alongside a stack of post-its with comparable thickness inside a standard paper envelope. In the second phantom, the same plastic box containing a different explosive (TNT) was placed alongside other objects inside a thicker cardboard box. In particular, a highlighter pen and a makeup removal pad were placed near the explosive, to develop a procedure that allows to simultaneously discriminate the explosive from more than one surrounding material. For both phantoms non-threat materials with a pronounced microstructure were chosen in order to 1) provide an appreciable dark-field signal and, thus, 2) to provide a challenge for discrimination of threat vs. non-threat materials.

As the quantitative parameter to determine the degree of material discrimination, we used the *CNR*, defined as:

$$CNR = \frac{|mean(ROI_1) - mean(ROI_2)|}{\sqrt{stdv(ROI_1)^2 + stdv(ROI_2)^2}} \qquad (1)$$

$ROI_{1,2}$ indicate Regions-Of-Interest selected inside the threat and non-threat material, respectively. The module at the numerator guarantees that the *CNR* is a positive value, and *stdv* indicates the standard deviation. The availability of five different contrast channels means that for a given set of 2 materials five different *CNR*s are available. We introduce the the linear combination of individual contrast channels in order to provide an integration of all contrasts into a single image:

$$I = a_1 I_{Abs_H} + a_2 I_{Abs_L} + a_3 I_{Offset} + a_4 I_{Scatt_H} + a_5 I_{Scatt_L} \quad , \quad (2)$$

where $a_{1-5}$ are free coefficients, and the pedices *$Abs_H$, $Abs_L$, Offset, $Scatt_H$, $Scatt_L$* refer to the intensities detected in the corresponding images. The *CNR* between two materials is then calculated while iterating over $a_{1-5}$ for the 5-contrast case, and over $a_{1-2}$ only for the dual energy "surrogate", and the set of coefficients resulting in the highest *CNR* value is selected. When only two materials need to be discriminated (first phantom), the above procedure is straightforward. When a certain target material (in our case the explosive) needs to be discriminated against more than one material (e.g. two, as in our second phantom), a two-step process is required. First, we calculated the minimum CNR between the material pairs for given set of coefficients $a_{1-5}$, which aims at the discrimination of threat materials from *all* non-threat materials. Second, we then iterate over coefficients $a_{1-5}$, and choose the set of coefficients that maximises the minimum CNR. This simultaneously maximises the distance (in *CNR* terms) between all three materials, which accounts for the possibility that the *ROI* corresponding to the material of interest is not known *a priori*.

**Results and Discussion**
The initial step was the calibration of the detector's higher threshold (Fig 2), essential to the ensuing developments.

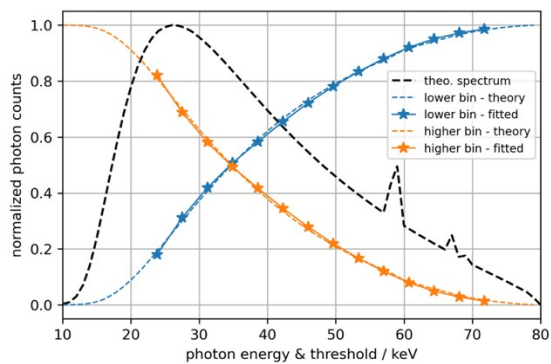

*Fig 2 calibration of the higher detector threshold*

The dashed black line in Fig. 2 represents the (normalised) 80 kVp spectrum of a tungsten source, obtained through SpekCalc [28-30]. Dashed orange and blue lines represent the total number of photons in the higher and lower energy bins (respectively) as a function of the threshold in keV, calculated from the same theoretical spectrum. Finally, the solid orange and blue lines with markers represent the experimentally detected number of counts as the real detector threshold (in V) is increased. As can be seen the curve shapes match very well, therefore determining the V/keV relation that leads to optimised overlap between experimental and theoretical curves results in a reliable threshold calibration. The calibration is given as $E_{th}[keV] = 3.69\, U_{th}[V] + 7.2$ with $E_{th}$ the energy threshold in keV and $U_{th}$ the threshold value in V.

The second preliminary step consisted in an outline determination of the threshold value that leads to an optimal *CNR* maximisation process. While this is in principle impossible to know *a priori* as it requires knowledge of the specific contrast values produced by the various materials in the different imaging channels, some degree of optimisation can be conducted on the basis of the background noise minimisation method based on a previous approach by Modregger *et al* [31]. The first step in this process is the determination of the system's sensitivity function, which in EI is the *illumination curve* (IC), obtained by scanning the pre-sample mask in the absence of a sample while the remainder of the imaging system is kept stationary [32]. This is modelled as a convolution between the (re-scaled) source distribution and the apertures in the pre-sample and detector masks, while taking into account a degree of transmission through the masks that gives rise to the IC's offset (Fig. 3a).

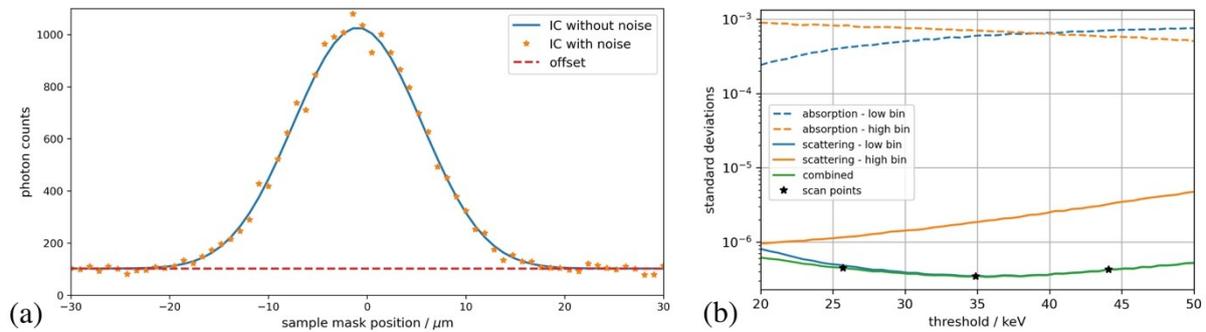

*Fig 3 modelling of the system's "illumination curve" (a) and noise behaviour in the various retrieved contrast channels (b). The green curve in the latter represents their combination via quadrature sum, and the black stars show the threshold values at which experimental images were collected in this study.*

The Modregger *et al* method then calculates the *stdv* in the background by retrieving "air" scans using the same process used in the retrieval of real images. Fig. 3(b) shows the *stdv* in the various contrast channels as a function of the energy threshold in KeV; their combination as a quadrature sum (green curve) provides an indication of the threshold value for which the combined *stdv* arising from all contrast values is minimised. As can be seen, the process is dominated by *Scatt$_L$*, and presents a broad minimum around 35 keV. While developing this process we observed that, despite the nominal gold thickness in the masks being 200 μm, a better match with the experimentally observed offset value is obtained with a gold thickness of 150 μm. This is not unprecedented, as masks are often affected by some degree of underplating as well as a reduced density compared to solid gold's nominal value [33]; however, this can be difficult to determine precisely as other factors (source tails, air scattering) can affect the offset value. For this reason, the above process was repeated using a gold thickness of 200 μm, with the results reported in the supplementary materials (suppl. fig. 1). As can be seen from that figure, the overall trend is very similar, with possibly a slight shift of the "optimal" threshold towards higher values; however, the broadness of the maximum and the indicative nature of the exercise (since, as mentioned, real contrasts are unknown *a priori*) means very similar indications are obtained in the two cases. However, to take this into account, the *CNR* optimisation process was repeated at three different threshold values, indicated with black stars in Fig. 3(b) and suppl. Fig. 1.

Fig. 4 presents the five retrieved images for the simpler, "two-material" phantom acquired with a detector threshold of 35 keV, which roughly matches the expected optimal noise behaviour as observed in Fig. 3(b). The second column in the table at the bottom right corner provides the "natural" contrast between post-its and Semtex in the various contrast channels. For completeness, the same table reports also the *CNR* of the two materials against the background (the envelope), although this has not been used in further analysis.

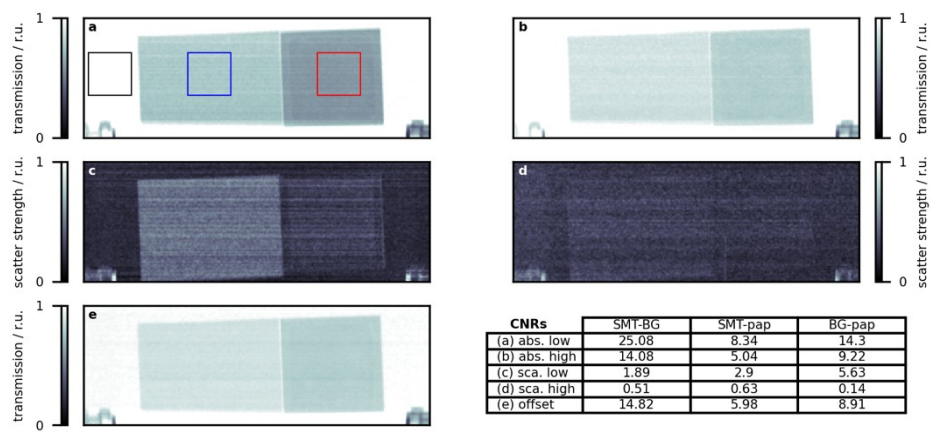

Fig 4 retrieved $Abs_L$ (a), $Abs_H$ (b), $Scatt_L$ (c) $Scatt_H$ (d) and Offest (e) images for the "two material" phantom acquired with a detector threshold of 35 keV. Paper (post-its) and Semtex 1H are visible on the left and right-hand sides, respectively. The ROIs from which mean and stdv values have been extracted for CNR calculation are shown in (a), with blue and red corresponding to paper and Semtex, respectively. The contrast against the background ("BG", black ROI) has also been calculated for completeness, although it has not been used for further calculations. The CNR in each image for each pair of materials is reported in the table at the bottom right corner.

Fig 5 reports the result of the "optimised linear contrast combination" applied to the above dataset, for two (top row) and five (bottom row) contrasts.

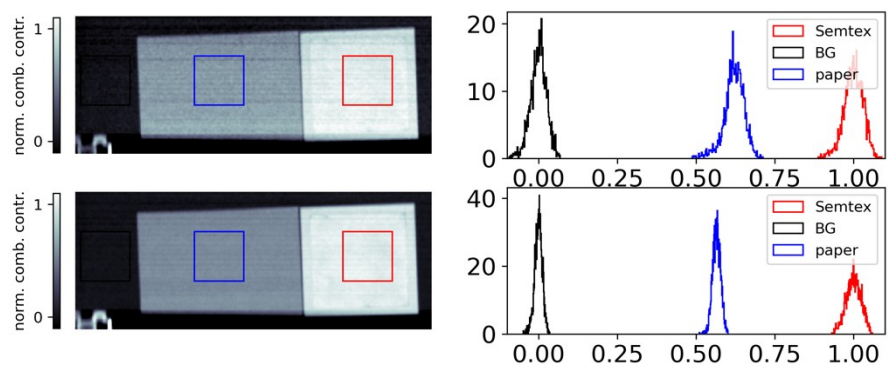

Fig 5 CNR optimisation at a 35 keV threshold for two ($Abs_L$ and $Abs_H$, top row) and five (bottom row) contrasts. Corresponding histograms extracted from the selected ROIs are shown at the right-hand side of each image (BG = background, i.e. the envelope).

The improvement in *CNR* is best appreciated by looking at the histograms on the right-hand side of each image, from which it is immediately evident that combining five contrasts makes the histograms much narrower and therefore the materials more neatly separated from each other. Optimised Semtex-paper (i.e. post-its) *CNR* values are 8.7 and 16.3 for the combination of two and five contrast respectively, indicating an almost 100% improvement resulting from the use of the three additional contrasts. By comparing this with the values reported in the second column at the bottom right corner of Fig 4, it can be noticed that the combination of $Abs_L$ and $Abs_H$ alone leads to a very small improvement over the attenuation values used on their own. We attribute the mere small improvement of combining the standard dual-energy contrasts to the fact that the noise between the $Abs_L$ and $Abs_H$ contrast channels was correlated (r=0.42), which can be explained by a redistribution of some photons from the high energy bin (i.e., $Abs_H$) to the low energy bin (i.e., $Abs_L$) by charge sharing [34]. Combining five contrasts, on the other hand, significantly outperforms all "native" values.

This exercise was repeated for detector thresholds of 26 and 44 keV, resulting in optimised *CNR* values of 7.7 and 9.3 (respectively) in the "two contrast" case, and of 15.1 and 15.4 in the "five contrast" case. This seems to indicate that the identification of 35 keV as the optimal threshold for the combination of five contrasts (Fig 3(b)) holds in the five-contrast case, although differences are small, as can be expected from the broadness of the minimum in the combined noise plot. The same does not apply to the two-contrast case (for which a 44 keV threshold gives a slightly higher value), however this could be expected as the combination of the noise levels was dominated by $Scatt_L$. Overall, the small differences among the above values indicates that a reasonable choice of threshold that roughly splits the spectrum in half provides close to optimal values, again in line with the broadness of the minimum observed in Fig. 3(b). The full datasets at thresholds 26 and 44 keV and their processed versions are reported for completeness in the supplementary materials, suppl. fig. 2 to 5.

Finally Fig. 6 shows the five retrieved images for the more complex phantom, still with a detector threshold of 35 keV.

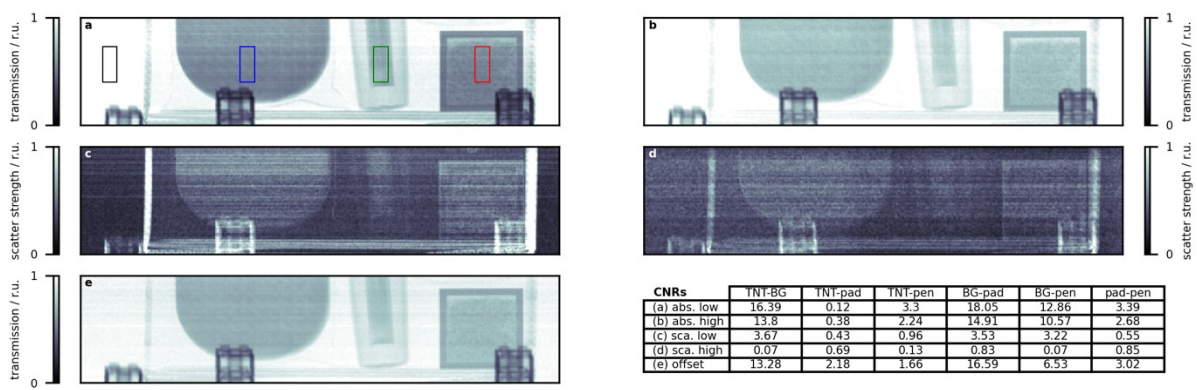

*Fig. 6 retrieved $Abs_L$ (a), $Abs_H$ (b), $Scatt_L$ (c) $Scatt_H$ (d) and Offset (e) images for the "three material" phantom acquired with a detector threshold of 35 keV. Pad, highlighter pen and TNT are visible from left to right in the images. The ROIs from which mean and stdv values have been extracted for the CNR calculations are shown in (a), with blue, green and red corresponding to pad, pen and TNT, respectively. The contrast between TNT and background ("BG", black ROI) has also been calculated for completeness, although it has not been used* *for further calculations. The CNR in each image for each pair of materials is summarised in the table at the bottom right corner.*

The columns of interest in the (more complex, due to the increased number of materials) table in the bottom right corner of Fig. 6 are the second and the third, indicating the "natural" *CNR* of TNT against pad and pen, respectively. As can be seen, in this case we are dealing with significantly lower contrasts than in the more simplistic case of two materials only, which we are aiming to enhance through the five-contrast combination process, the results from which are shown in Fig. 7. The optimisation on the dual contrast dataset (*Abs$_L$* and *Abs$_H$*, top row) gives a *CNR* of 0.7 for both TNT vs pen and TNT vs pad, which is a gain (0.7 vs 0.4) in the TNT vs pad case but a loss (0.7 vs 3.3) in the TNT vs pen case. This can be expected, since the algorithm simultaneously maximises the relative distance between all material pairs, which is the only possible approach on the assumption that the target material is unknown. This notwithstanding, when all five contrast are used, a *CNR* of 3.7 is obtained for both TNT vs pen and TNT vs pad, which is higher than all native *CNR* values (3.7 vs 2.2 for TNT vs pad and 3.7 vs 3.3 for TNT vs pen), when the "best of all five" is selected for the latter. Clearly with some degree of prior information being available (e.g. contrast boundaries for the material of interest obtained through previous calibration), the algorithm performance could be significantly improved.

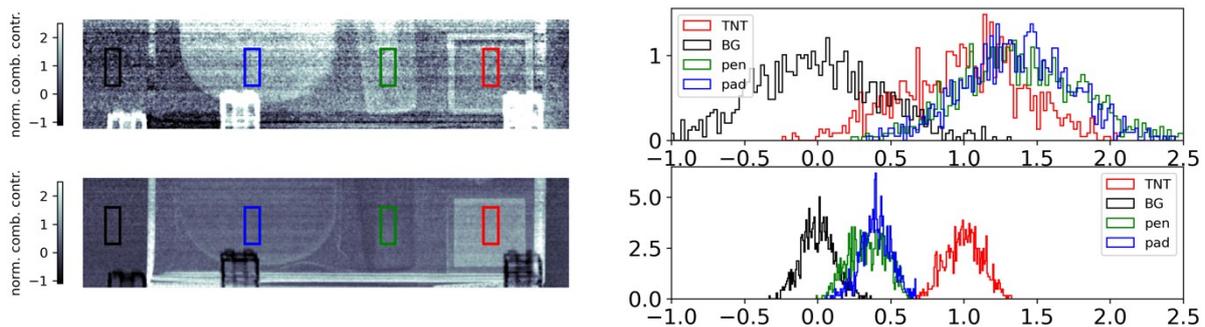

*Fig 7 CNR optimisation at a 35 keV threshold for two (Abs$_L$ and Abs$_H$, top row) and five (bottom row) contrasts. Corresponding histograms extracted from the selected ROIs are shown at the right-hand side of each image, and labelled accordingly (BG = background, i.e. the cardboard box). The narrowing of the histograms resulting from the combination of five contrasts is even more evident than in the previous example.*

Also in this case results at detector thresholds of 26 and 44 keV are reported for completeness in the supplementary materials (Suppl. Fig. 6 to 9). At 26 keV, optimal *CNR*s of 0.9 and 1.9 are obtained for TNT vs pad and TNT vs pen with two contrasts, versus maximum native values of 0.7 and 2.6. For the five-contrast combination, the *CNR* becomes 2.1 for TNT vs both materials, leading to a gain in the pad case (vs 1.8) but a loss in the pen case (vs 2.6, as reported above). At 44 keV, an optimal *CNR* of 1.6 is obtained for TNT vs both materials with two contrast, versus maximum native values of 1.2 and 1.5, leading to a small gain in both cases. With five contrasts, however, the gain is more significant with a *CNR* of 4.3 for TNT vs both materials, vs native maxima of 1.6 (TNT-pad) and 3.4 (TNT-pen). This is an even greater gain than observed at 35 keV, which supports the trend observed in Suppl. Fig. 1 in which higher threshold values seem to be slightly more advantageous. While this would seem to support the assumption of a slightly thicker gold layer in the masks, it should also be noted that the simplistic model based purely on noise behaviour we used to obtain fig. 3(b) and Suppl. Fig. 1 may be insufficient to describe the increasingly complex case where multiple materials are present and their respective *CNR*s need to be simultaneously maximised.

**Conclusions**
This paper provides proof-of-concept evidence that the inclusion of additional contrast mechanisms in an imaging system can aid the discrimination between materials with similar attenuation characteristics. The study is admittedly preliminary, and used the optimisation of a simple linear combination of two and five contrast to maximise the *CNR* between material pairs, and demonstrate the increased detectability that can be provided by the inclusion of additional contrast channels. While the approach is straightforward when applied to material pairs, the inclusion of additional materials leads to an increased degree of compexity, mostly related to the need to maximise the *CNR* between each material pair when no *a priori* information is available on the target material. However, even such a simple framework is sufficient to prove that room for improvement exists, which we hope will trigger further research in this direction.

**Acknowledgments**
This work was supported by the EPSRC (Grant EP/T005408/1). Additional support was obtained through the Innovative Research Call in Explosives and Weapons Detection 2016. This is a Cross-Government programme sponsored by a number of Departments and Agencies under the UK Government's CONTEST strategy in partnership with the US Department of Homeland Security, Science and Technology Directorate. AO was supported by the Royal Academy of Engineering under their Chairs in Emerging Technologies scheme.

**Declaration of Interest**
AA, IGH and DB are, or were at the time the research was carried out, Nikon employees. AO is a named inventor on patents owned by UCL protecting the technology used to obtain the described results. PM has no conflicts of interest to disclose.

**Supplementary materials** to **Increased material differentiation through multi-contrast x-ray imaging: a preliminary evaluation of potential applications to the detection of threat materials**

A. Astolfo[1,2], I.G. Haig[2], D. Bate[2,1], A. Olivo[1], P. Modregger[3,4]

1. Department of Medical Physics and Biomedical Engineering, UCL, London WC1E 6BT, UK
2. Nikon X-Tek Systems Ltd, Tring, Herts, HP23 4JX, UK
3. Department of Physics, University of Siegen, 57072 Siegen, Germany
4. Center for X-ray and Nano Science CXNS, Deutsches Elektronen-Synchrotron DESY, 22607 Hamburg, Germany

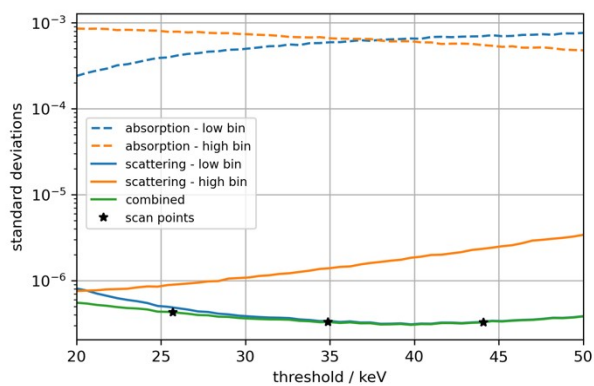

**Suppl. Fig 1.** Noise behaviour in the various retrieved contrast channels as a function of the detector threshold, for a mask gold thickness of 200 μm. As can be seen, very little difference in the overall trend are observed compared to the 150 μm thickness case presented in the main article. The green curve provides an indication of the overall noise behaviour when contrasts are combined by taking their quadrature sum, and the black stars represent the threshold values that have been tested in the actual experiment.

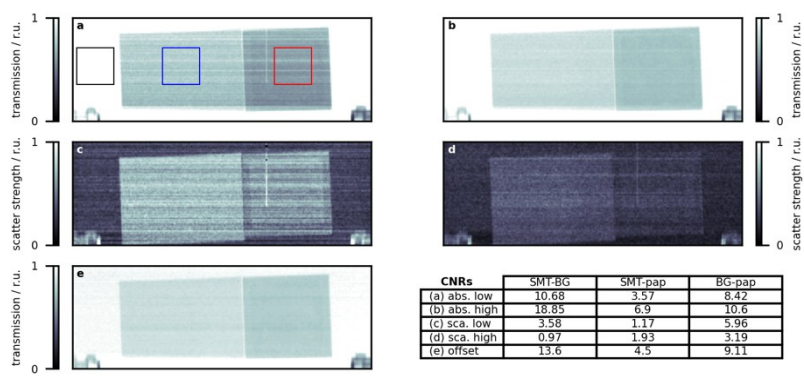

**Suppl. Fig. 2.** Retrieved $Abs_L$ (a), $Abs_H$ (b), $Scatt_L$ (c) $Scatt_H$ (d) and Offest (e) images for the "two material" phantom acquired with a detector threshold of 26 keV. Paper (post-its) and Semtex 1H are on the left and right-hand sides, respectively. The ROIs from which mean and *stdv* values have been extracted for *CNR* calculation are shown in (a), with blue and red corresponding to paper and Semtex, respectively. The contrast against the background ("BG", black ROI) has also been calculated for completeness. The *CNR* in each image for each pair of materials is summarised in the table at the bottom right corner.

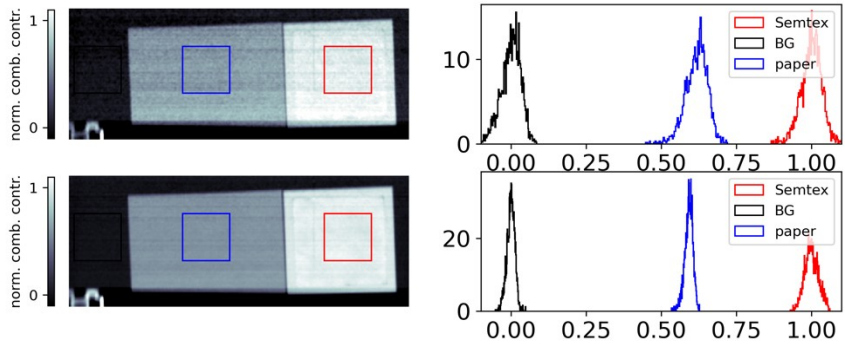

**Suppl. Fig. 3.** *CNR* optimisation at a 26 keV threshold for two ($Abs_L$ and $Abs_H$, top row) and five (bottom row) contrasts. Corresponding histograms extracted from the selected ROIs are shown at the right-hand side of each image (BG = background, i.e. the envelope).

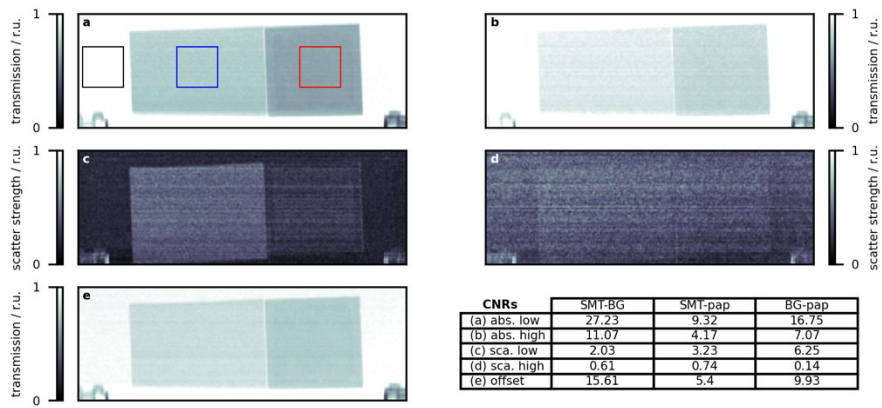

**Suppl. Fig. 4.** Retrieved $Abs_L$ (a), $Abs_H$ (b), $Scatt_L$ (c) $Scatt_H$ (d) and Offest (e) images for the "two material" phantom acquired with a detector threshold of 44 keV. Paper (post-its) and Semtex 1H are on the left and right-hand sides, respectively. The ROIs from which mean and *stdv* values have been extracted for *CNR* calculation are shown in (a), with blue and red corresponding to paper and Semtex, respectively. The contrast against the background ("BG", black ROI) has also been calculated for completeness. The *CNR* in each image for each pair of materials is summarised in the table at the bottom right corner.

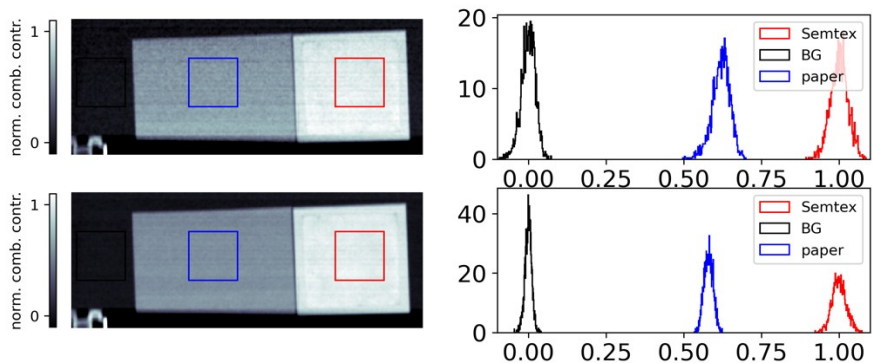

**Suppl. Fig. 5.** *CNR* optimisation at a 44 keV threshold for two ($Abs_L$ and $Abs_H$, top row) and five (bottom row) contrasts. Corresponding histograms extracted from the selected ROIs are shown at the right-hand side of each image (BG = background, i.e. the envelope).

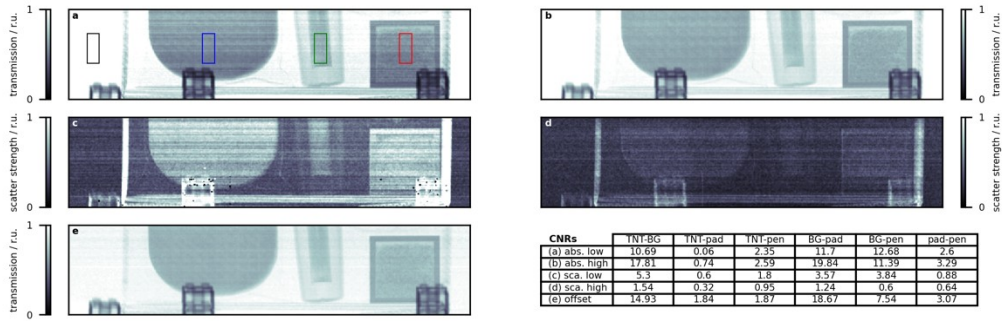

**Suppl. Fig. 6.** Retrieved $Abs_L$ (a), $Abs_H$ (b), $Scatt_L$ (c) $Scatt_H$ (d) and Offest (e) images for the "three material" phantom acquired with a detector threshold of 26 keV. Pad, highlighter pen and TNT are visible from left to right. The ROIs from which mean and stdv values have been extracted for *CNR* calculation are shown in (a), with blue, green and red corresponding to pad, pen and TNT, respectively. The contrast between TNT and background ("BG", black ROI) has also been calculated for completeness, although it has not been used for further calculations. The *CNR* in each image for each pair of materials is summarised in the table at the bottom right corner.

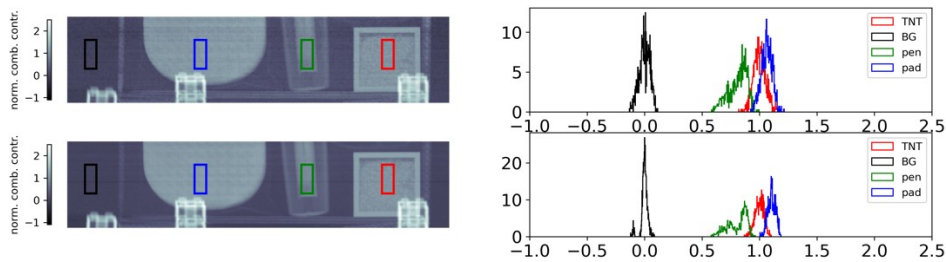

**Suppl. Fig. 7.** *CNR* optimisation at a 35 keV threshold for two ($Abs_L$ and $Abs_H$, top row) and five (bottom row) contrasts. Corresponding histograms extracted from the selected ROIs are shown at the right-hand side of each image, and labelled accordingly (BG = background, i.e. the cardboard box).

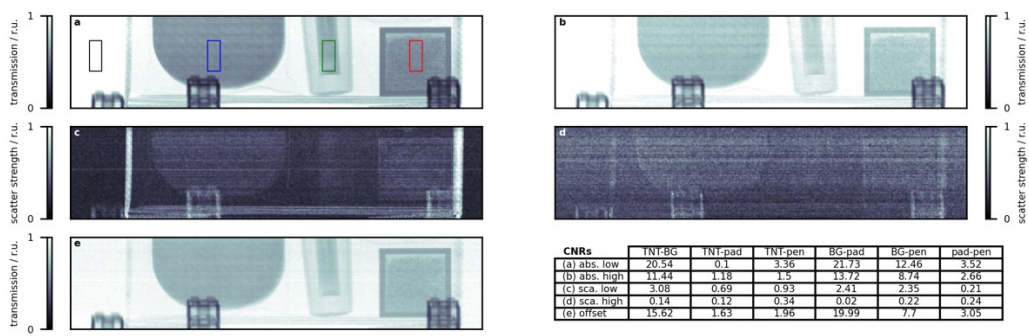

**Suppl. Fig. 8.** Retrieved $Abs_L$ (a), $Abs_H$ (b), $Scatt_L$ (c) $Scatt_H$ (d) and Offest (e) images for the "three material" phantom acquired with a detector threshold of 44 keV. Pad, highlighter pen and TNT are visible from left to right. The ROIs from which mean and stdv values have been extracted for *CNR* calculation are shown in (a), with blue, green and red corresponding to pad, pen and TNT, respectively. The contrast between TNT and background ("BG", black ROI) has also been calculated for completeness, although it has not been used for further calculations. The *CNR* in each image for each pair of materials is summarised in the table at the bottom right corner.

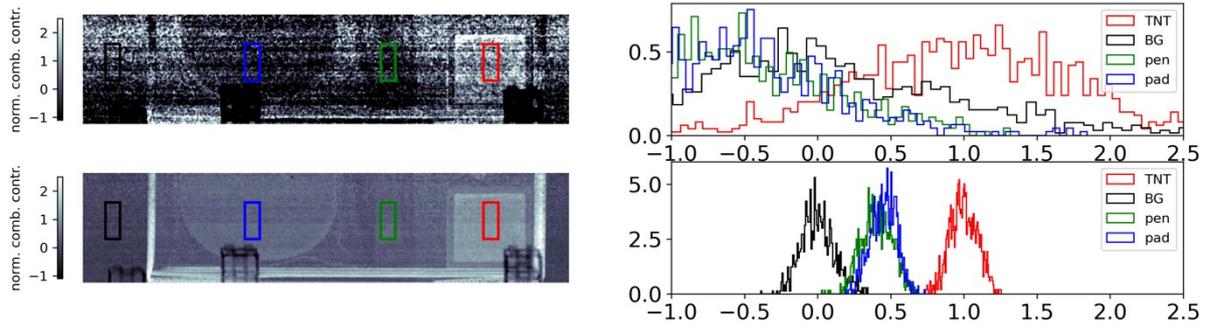

**Suppl. Fig. 9.** *CNR* optimisation at a 44 keV threshold for two ($Abs_L$ and $Abs_H$, top row) and five (bottom row) contrasts. Corresponding histograms extracted from the selected ROIs are shown at the right-hand side of each image, and labelled accordingly (BG = background, i.e. the cardboard box).